\documentclass[a4paper,superscriptaddress,twocolumn,preprintnumbers,showpacs,amsmath,amssymb,prl]{revtex4-1}
\usepackage{graphicx}
\usepackage{latexsym}
\usepackage{mathrsfs} 

\begin{document}
\title{Peptide pores in lipid bilayers : voltage facilitation pleads for a revised model}

\author{G.C. Fadda}
\affiliation{Laboratoire L\'eon Brillouin, CEA/CNRS UMR 12, CEA-Saclay, 91191 Gif-sur-Yvette, France}
\affiliation{Universit\'e Paris 13, UFR SMBH, 74 rue Marcel Cachin, 93017 Bobigny, France}

\author{D. Lairez}
\thanks{Corresponding author}
\email{lairez@cea.fr}
\affiliation{Laboratoire L\'eon Brillouin, CEA/CNRS UMR 12, CEA-Saclay, 91191 Gif-sur-Yvette, France}

\author{Z. Guennouni}
\affiliation{Laboratoire L\'eon Brillouin, CEA/CNRS UMR 12, CEA-Saclay, 91191 Gif-sur-Yvette, France}
\affiliation{Institut des nanoSciences de Paris, UMR 7588, UPMC, 75005 Paris, France}

\author{A. Koutsioubas}
\altaffiliation[Present address~: ]{JCNS at FRM II, Forschungszentrum J\"ulich GmbH, Lichtenbergstr. 1, 85747 Garching, Germany}
\affiliation{Laboratoire L\'eon Brillouin, CEA/CNRS UMR 12, CEA-Saclay, 91191 Gif-sur-Yvette, France}

\date{\today}
\preprint{accepted for publication in \textit{Phys. Rev. Lett.}}

\begin{abstract}
We address the problem of antimicrobial peptides that create pores in lipid bilayers, focusing on voltage-temperature dependence of pore opening. Two novel experiments (voltage-clamp with alamethicin as an emblematic representative of these peptides and neutron reflectivity of lipid-monolayer at solid/water interface under electric field) serve to revise the only current theoretical model\,\cite{Huang:2004fk}. We introduce a general contribution of peptide adsorption and electric field as being responsible for an unbalanced tension of the two bilayer leaflets and we claim that the main entropy cost of one pore opening is due to the corresponding "excluded-area" for lipid translation.
\end{abstract}

\maketitle

Interaction of living cell membrane with adsorbed molecules and the way their uptake occurs are at the heart of many biological issues. Among these molecules, antimicrobial peptides\,\cite{Zasloff:2002fk, Bechinger:1997} attract special attention as being the keystone of the innate immune system of multicellular organisms. Antimicrobial peptides basically cause lipid-bilayer permeation by producing pores. Their universal presence in animal and plant kingdoms, their non-specific broad-spectrum and their elementary structure let us expect their action also obeys to a widespread and universal physical mechanism that probably puts the membrane behavior in central position.

The most accepted physical model for peptide pores opening is based on the following tension-driven mechanism\,\cite{Huang:2004fk}. Prior to form pores, these amphipathic peptides adsorb parallel onto the membrane and are supposed to increase an "internal tension" up to a given adsorption level, beyond which they relax this tension by penetrating into the membrane, then stabilize the edge of pores that spontaneously appear in bilayers\,\cite{Taupin:1975fk}. Although this model is  the best attempt to formalize a widespread outlook, it is still unsatisfactory as it ignores two points~: 1)~the role of temperature and entropy~; 2)~the role of the transverse electric field of the order of $25\times 10^6$\,V/m experienced by living cell membranes that is known to be strong enough in some cases\,\cite{savko:1982} to induce peptide-pores.

Here, we report two novel experiments~: 
1)~voltage-clamp focusing on the temperature-voltage dependence of pore opening with alamethicin as an emblematic representative of antimicrobial peptides\,\cite{Cafiso:1994}~;
2)~neutron reflectivity of lipid-monolayer adsorbed at solid/water interface under electric field. We show that the membrane behavior is central and the key role of its entropy. 
We support the idea that voltage-induced and zero-voltage peptide-poration obey the same physics~: the former guiding us to propose a simple model that takes over some fundamentals of the tension-driven mechanism but solves some difficulties. In particular, we clarify the above "internal tension" as being due to  a bending energy rather than a proper membrane tension~; we introduce a general contribution of peptide adsorption and electric field as being responsible for an unbalanced tension of the two bilayer leaflets~; and finally we show that for a held membrane the main entropy cost of one pore opening comes from the corresponding excluded-area for the translational entropy of lipids. 

\textit{Voltage-clamp~:}
Pore opening was detected by measuring the ionic current as an electric potential is applied across a free-standing planar membrane between two compartments containing KCl-1M aqueous solutions (for sample preparation and experiment setup see ref.\,\cite{Fadda:2009}). Here, alamethicin-poration was studied on DPPC-bilayer versus temperature above the melting point of the lipid.
This was checked by measuring at 10\,Hz, resistance and capacitance of the bilayer as a function of temperature. Both are constant above 299\,K. The measured specific capacitance $ 0.5$\,$\mu$F/cm$^2$ is consistent with usual values\,\cite{White:1970fk}.

\begin{figure}[!htbp]
\centering
\includegraphics[width=0.8\linewidth]{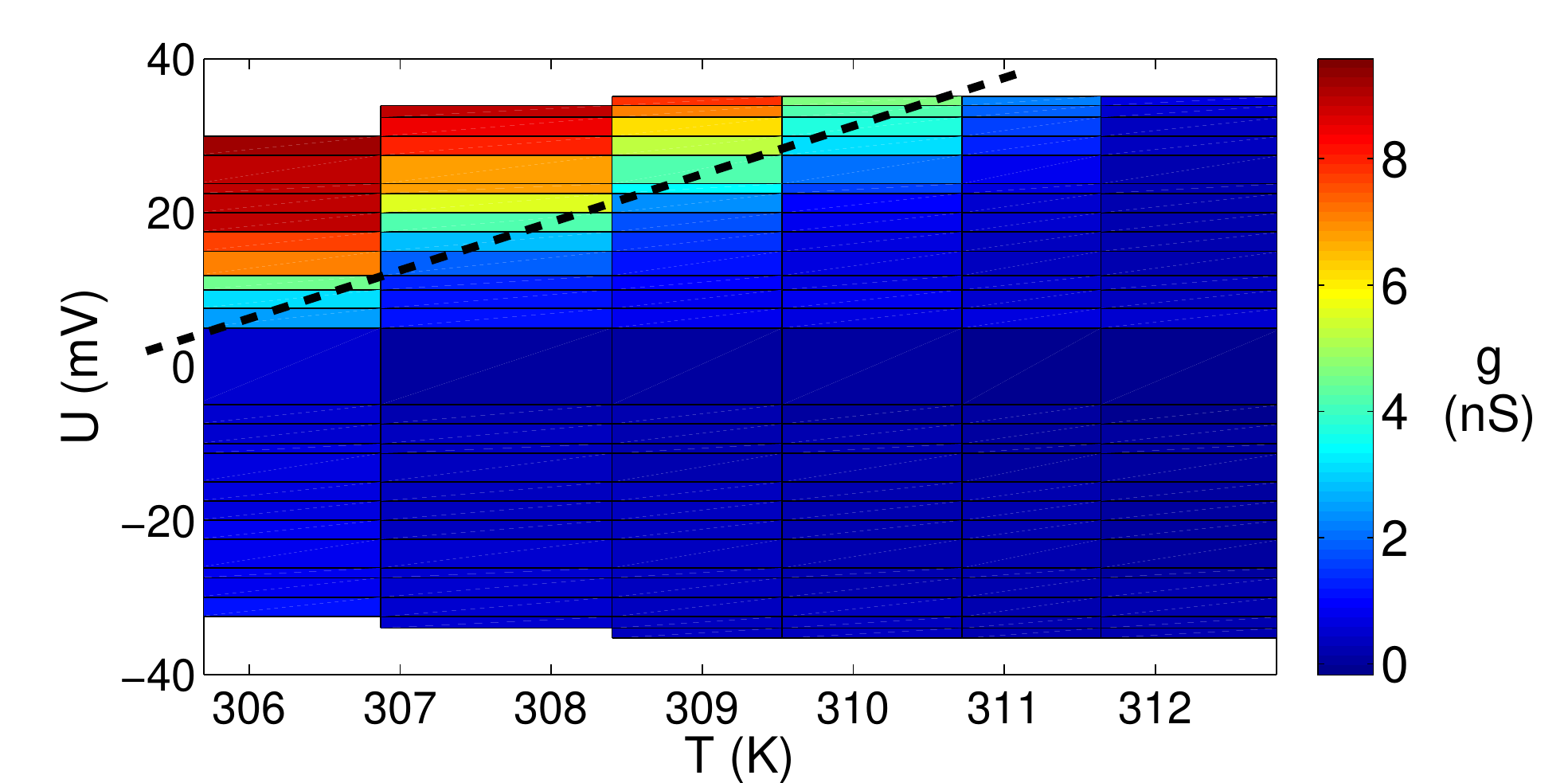}
\caption{Conductance $g$ of DPPC/alamethicin vs. temperature $T$ and electric potential $U$ of the alamethicin compartment. The dashed-line has a slope equal to 6\,mV/K.}
\label{fig3}
\end{figure}

In our study, alamethicin was added only to one side of the membrane (\textit{cis}-side) at a molecular ratio lipid/peptide of the order of 100 (estimated from the area to volume ratio of the device, the membrane area and the amount of peptides).
At a given temperature $T$, the current intensity $I$ was recorded while the voltage $U$ was alternatively set to positive and negative values of increasing modulus (polarity refers to \textit{cis}-side). For sections of positive voltage above a given threshold, the current has a significant non-zero value, but remains almost zero for negative voltage. This reveals the formation of pores, which are induced by electric field with the proper \textit{cis-trans} direction, and are removed upon field inversion. Fluctuations of current have been already analyzed\,\cite{Fadda:2009}. Here, we focus on the average conductance $g=I/U$ computed for each voltage-section as a function of $T$ and $U$. Fig.\ref{fig3} shows a typical result. At this ionic strength the typical conductance of a single pore is 1\,nS~\cite{Fadda:2009}, so the dashed-line in Fig.\ref{fig3} might correspond to the peptide-poration. Note that its temperature dependence is the opposite of thermally activated processes such as electroporation without peptide. Also, the asymmetry regarding polarity is absent for electroporation and can only be explained by the asymmetric peptide addition. This supports that the dashed-line in Fig.\ref{fig3} coincides with the peptide-poration transition. Note that this interpretation is also fully consistent with circular-dichroism experiments that have shown that the association of alamethicin decreases with temperature\,\cite{Woolley:1993uq}.

The temperature dependence of the transition reveals the role of entropy, whereas the voltage variation shows an electric energy term. With the pore-free state as reference, the Gibbs free energy per pore is 
$G=H + QU - TS$, with $H$ the enthalpy gain for one pore opening, $S$ and $QU$ the entropy loss and electric work, respectively. At equilibrium $H + QU^* = T^*S$, leading to~: ${dU^*}/{dT^*}={S}/{Q} $.
Here we found $S/Q\simeq 6$\,mV/K, or:
\begin{equation}\label{S}
{TS}/{kT}\simeq 70 \times Q/\text{e}
\end{equation}
with $k$ the Boltzmann constant and $\text{e}$ the electron charge.

Due to its rodlike $\alpha$-helix structure, alamethicin has a global permanent electric dipole $\mu_{\text{P}}=15$\,$\text{e\AA}$\,\cite{savko:1982}, which could be responsible for voltage effects\,\cite{Bechinger:1997}. This idea supposes a parallel (rather than antiparallel) orientation of peptides forming a pore. If this were correct, assuming 6 peptides per pore\,\cite{Qian:2008ve} gives $Q=6\mu_{\text{P}}/z$. A typical value for the bilayer thickness $z\simeq 45\text{\AA}$\,\cite{naggle:2000} and Eq.\ref{S} would lead to ${TS}\simeq 140\text{\,}kT$. This is nonphysical and in turn pleads for an antiparallel orientation of dipoles so that their moments cancel. This is not so amazing since~: 
1)~aligning one peptide dipole to the field would save  $\sim1\text{\,kT}$, but it would cost much more by confining parallel dipoles in repulsive interaction to form a pore~; 
2)~the peptide dipole cannot explain the asymmetry regarding polarity as it can always favorably align with the field~; 
3)~some peptides form voltage-facilitated pores despite them having no dipole\,\cite{Sokolov:1999fk}. So, the driving force for voltage-induced peptide-pores necessarily originates from the membrane.

\textit{Neutron reflectivity~:}
Little is known about lipid bilayers under electric field. Structural effects were observed by infrared spectroscopy\,\cite{Zawisza:2007fk} or neutron reflectivity\,\cite{Burgess:2004} on dried stacked assemblies of bilayers. For high electric field ($\gtrsim 10^8$\,V/m) results  suggest the alignment with the field of phospholipid zwiterionic head-groups. 
Could similar effects occur at lower field (i.e. comparable to the natural transmembrane field) for fully hydrated head-groups\,? Neutron reflectivity experiments have been reported on floating bilayers near a solid-water interface\,\cite{Lecuyer:2006vn}. On such membranes, undulations that increase in amplitude with electric field, dominate the reflectivity spectrum and likely hide more subtle changes.

To overcome this issue, we performed neutron reflectivity (EROS/LLB) on single monolayers adsorbed at the interface between saline heavy water ([KCl]=1\,M) and conductive silicon wafer  allowing the electric field to be applied\,\cite{Koutsioubas:2012a}. Si-wafers were silanized following ref.\,\cite{Sanjuan:2007} using octadecylthrichlorosilane (OTS). 
DPPC-monolayer was deposited on silanized wafer by a modified Langmuir-Shaefer technique avoiding contact of the deposited layer with air. Deposits were done at $25^{\circ}\text{C}$ in the "liquid-expanded" (LE) phase at the controlled number density of $1/A_\text{L}=1/86$\AA$^2$ comparable to that of aqueous bilayers and ensuring the fluidity of the film.
The overall capacitance was measured as $C=1\text{\,}\mu\text{F}$. 
Let us define the average electric field in the monolayer as $E=(C/A\epsilon_m)U$, with $\epsilon_m=3\epsilon_0$ the effective membrane permittivity\,\cite{Stern:2003}, $\epsilon_0$ that of vacuum and $A=12.6\text{\,cm}^2$ the area.
One gets~: $E=30\times10^6\text{\,V/m}$ for $U=1\text{\,V}$ that is comparable to natural transmembrane field%
\,\footnote{Here, the crude approximation that the bilayer is homogeneous with an effective permittivity $\simeq 3$, just aims to compare the electrical field in our experiment to similar estimations of the natural transmembrane field.}.
\begin{figure}[!htbp]
\centering
\includegraphics[width=1\linewidth]{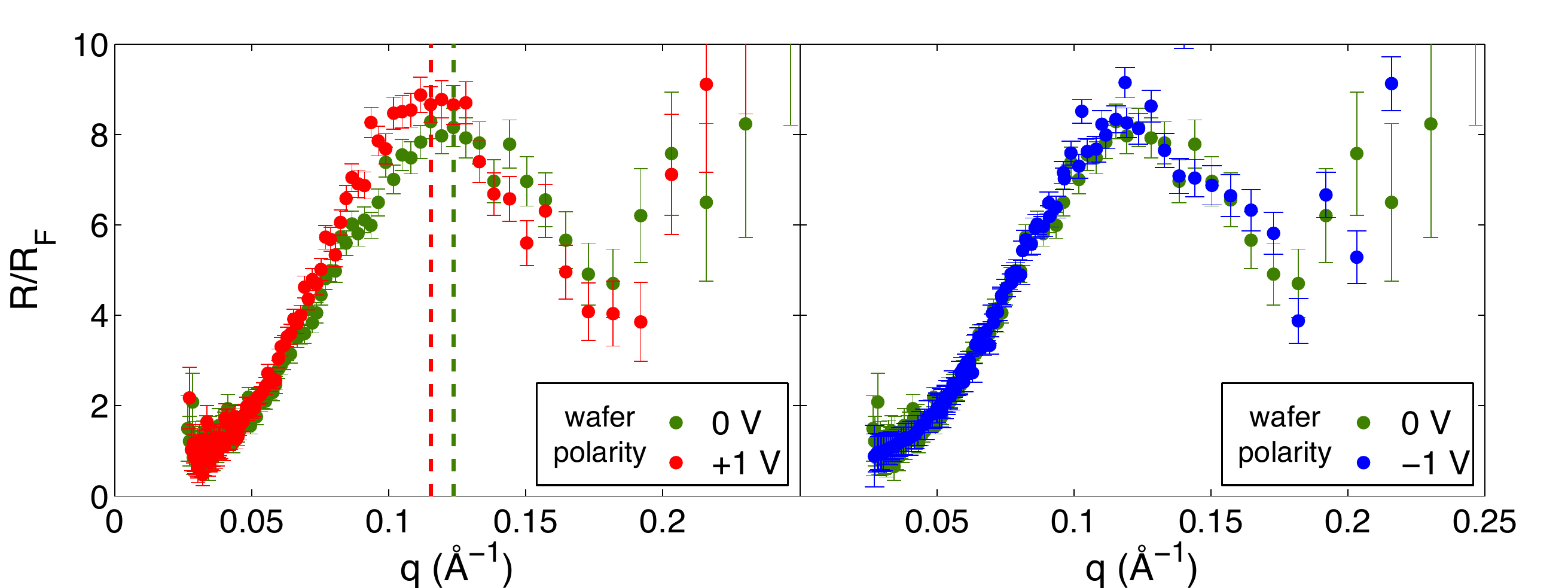}
\caption{Reflectivity of DPPC-monolayer at D$_2$O/OTS-Si-wafer interface divided by the Fresnel's reflectivity of D$_2$O/Si vs. transfer vector $q$.
Dashed lines mark oscillations maxima.}
\label{figref}
\end{figure}

Reflectivity was measured at $23^\circ\text{C}$ for wafer potential $U=0, -1, +1\text{\,V}$. Measurements reliability was checked at least three times. We always observed that spectra divided by the Fresnel reflectivity show characteristic oscillations, which are fully superimposable for $U=0$ and $-1\text{\,V}$, but are unambiguously shifted to low values of transfer vector $q$ for $+1\text{\,V}$  (Fig.\ref{figref}). This shift is irreversible at this temperature. In a quite general manner, it corresponds to thickening of the overall layer by~:
\begin{equation}\label{deltaz}
\Delta h\simeq\left({1/q_{\text{+1V}}^{*}-1/q_{\text{0V}}^{*}}\right)\times2\pi= (3.6\pm0.2)\text{\,\AA}
\end{equation}
with $q^*$ the positions of first maximum\,%
\footnote{Calculation of reflectance involves 4 layers (silicon oxide, silane, lipid tails and heads) with 2 or 3 parameters each (density, thickness, roughness). Fitting data with these 8 or 12 free parameters would result in large confidence intervals, which finally for our purpose do not provide more informations than Eq.\ref{deltaz} does.}. 
This thickening is the opposite of expectation for capacitive compression. The asymmetry regarding polarity is the major result of this section, it necessarily involves the only permanent dipoles, i.e. those of head-group zwitterions of moment $\mu_{\text{L}}=4\text{\,e\AA}$\,\cite{SHEPHERD:1978uq}. Given the size of head-groups\,\cite{naggle:2000}, Eq.\ref{deltaz} is fully consistent with their orientation almost parallel to the plane at 0 or -1\,V and aligned to the field for +1\,V.
Due to incompressibility, thickening goes with the area shrinking~: $\Delta A_\text{L}/A_\text{L}=-\Delta h/h$. With $h=z/2\simeq 22.5$\,\AA\,\cite{naggle:2000}, one gets $\Delta A_\text{L}\simeq -14$\AA$^2$ that may cause on a solid substrate the transition from LE to a more condensed phase and could explain the observed irreversibility.
For a free bilayer the area shrinking only concerns the cathodic-leaflet and tends to bend the membrane with convexity on anode-side. 
For a held bilayer, the area shrinking is not real and only rises the tension of the cathodic-leaflet and the spontaneous curvature, consistently with molecular dynamics simulations\,\footnote{For bilayers and 3 times higher electric field, MD simulations show a 5$^\circ$ tilt of headgroups for the cathodic leaflet but no effect for the anodic leaflet (R. A. B$\ddot{\text{o}}$ckmann, B. L. de Groot, S. Kakorin, E. Neumann, and H. Grubm$\ddot{\text{u}}$ller, Biophys. J., \textbf{95}, 1837 (2008)), consistently with the asymmetry here reported and our interpretation}.

\textit{Model for peptide-poration}~-~%
Spontaneous opening of one pore in bilayer involves a mechanical energy\,\cite{Taupin:1975fk,Glaser:1988vn}~:
\begin{equation}\label{taupin}
E_m=\gamma_e 2\pi r  - \sigma_{m0} \pi r^2
\end{equation}
with $r$ the radius of the pore, $\gamma_e$ its edge-energy and $\sigma_{m0}$ the bare-membrane tension. $E_m$ reaches a maximum $E^*_m=\pi\gamma_e^2/\sigma_{m0}$ for $r_m^*=\gamma_e/\sigma_{m0}$ below which the pore tends to close and beyond which it grows indefinitely. 

\textit{1) Tension-driven mechanism}\,\cite{Huang:2004fk}~: In this now classical model, a peptide adsorbed onto a bilayer pushes away the lipid head-groups in order that its hydrophobic part meets the heart of the bilayer. Thus, a symmetric adsorption on both leaflets expands the area $A$ of the bilayer, accordingly with its thinning. For small area number density $x$ of peptides~: $\Delta A/A=xA_\text{P}/2$, with $A_\text{P}$ the area per peptide. This is supposed to equally increase the membrane tension $\sigma_m=k_s (\Delta A/A)$, with $k_s$ the stretching modulus~:
\begin{equation}\label{canonical}
{\partial\sigma_m}/{\partial x}=k_sA_\text{P}/2
\end{equation}
$\sigma_m$ would increase until $E^*_m$ is small enough to allow thermal fluctuations to form pores larger than $r_m^*$. Let us denote $x^*$ the corresponding $x$-value. Assuming that peptides incorporated into the membrane at the pore edge have no effect on surface tension, it can be shown that further increase of $x$ enriches the incorporated-peptide population only (i.e. their area number density $x_i$ increases whereas $x-x_i=x^*$ is constant), similarly to a phase transition.
Thus, beyond $x^*$ the contribution of adsorbed-peptides to $E_m$ in Eq.\ref{taupin} is $E_{\text{ap}}=- \pi r^2(x^*  k_sA_\text{P}/2)$. 
If the line density $\rho$ of peptides on the pore rim is constant~: $x_i=n_i\rho 2\pi r$, with $n_i$ the area number density of pores, then $E_{\text{ap}}$ splits into two terms~:
\begin{equation}\label{huang}
E_{\text{ap}}= - \left({x k_sA_\text{P}/2}\right)\pi r^2 + \left({2\pi^2n_i\rho k_sA_\text{P}/2}\right) r^3
\end{equation}
With Eq.\ref{taupin} in mind, the term in $r^3$ allows pores of radius larger than $r_m^*$ to be stable. This model faces two problems : 
1)~Amphipathic peptides have a positive surface excess (they populate the surface rather than the membrane bulk) and from the Gibbs adsorption isotherm $\partial\sigma_m/\partial x$ should be negative\,%
\footnote{In ref.\cite{Huang:2004fk}, the membrane is a canonical ensemble. Eq.\ref{canonical} should be revised as the grand canonical ensemble is likely to be more adequate}; 
2)~This model cannot account for voltage-induced-pores as the electric field causes a capacitive pressure on bilayer that lowers its tension\,\cite{Lacoste:2009}.

\textit{2) Revised model}~:
Based on our reflectivity results, we propose to solve the above issues by noting that when peptides are asymmetrically adsorbed on one given bilayer leaflet, they tend to increase its "natural" area and thus the spontaneous bilayer curvature, $c_0$, with the same convexity orientation as \textit{cis-trans} electric field does. This can be the origin of the "internal tension" of ref\,\cite{Huang:2004fk}.

\textit{- Spontaneous curvature}~: The curvature energy stored by a flat bilayer is $E_{el}=A\frac{1}{2}k_cc_0^2$, with $k_c$ the bending elastic modulus. For spherical curvature $c_0\ll z^{-1}$, $c_0=\Delta A/A2z$, with $\Delta A$ the area difference between leaflets. For $x$ peptides per unit area adsorbed on one given leaflet, we obtain~: $\Delta A/A=xA_\text{P}$. Thus $c_0=x\mathcal{L}$ and $E_{el}=A\frac{1}{2}k_c \mathcal{L}^2 x^2$, with $\mathcal{L}=A_\text{P}/2z$. On the contrary, peptides in pores equally contribute to both leaflets and do not affect $E_{el}$ (as for $\sigma_m$ within the tension-driven mechanism). So, with one pore involving $\rho2\pi r$ peptides, the curvature energy is reduced to $E_{el}=A\frac{1}{2}k_c \mathcal{L}^2 (x-\rho2\pi r/A)^2$. The energy difference is~:
\begin{equation}\label{Emp}
E_{\text{ap}}=-k_c\mathcal{L}^2 x \rho \times 2\pi r,
\end{equation}
that replaces Eq.\ref{huang} of the tension-driven mechanism\,\cite{Huang:2004fk}. Note that by establishing bridges between the two bilayer leaflets, pores increase their coupling that in turn increases the bending rigidity. This effect has been observed by neutron spin echo\,\cite{Lee:2010cr}. It can be the origin of cooperativity of pore opening.

\textit{- Pore edge}~:
Substituting Eq.\ref{huang} by Eq.\ref{Emp}, rises the issue of pore stability that we propose to solve here. Although antimicrobial peptides are often rodlike, their amphiphilicity has not the symmetry of a solid of revolution. Instead, its projection in the plane normal to the axis is bipolar\,\cite{tossi:2000}. Incorporated peptides minimize their interfacial energy $\mathcal{E}_{\text{P}}$ when they are at the pore edge with their hydrophilic zone facing the channel. Consider the channel-section as a polygon of angle $\alpha$ with peptides as vertices and denote $\alpha^*$ the angle of same vertex that includes the hydrophilic zone of the peptide ($\alpha^*<\pi$ otherwise hydrophilicity dominates and peptides likely do not adsorb). $\mathcal{E}_{\text{P}}$ is minimum for $\alpha=\alpha^*$ at the optimal pore radius $r^*=1/\left({\rho(\pi-\alpha^*)}\right)$. Deviation from $\alpha^*$ increases~$\mathcal{E}_{\text{P}}=r_{\text{P}}z\Delta\alpha\Delta\gamma$, with $r_{\text{P}}$ the radius of peptide-rod, $\Delta\alpha=\|\alpha-\alpha^*\|$, and $\Delta\gamma>0$ the surface tension difference between heterophilic and homophilic contacts ($\Delta\gamma_{\alpha<\alpha^*}=\gamma_w^h-\gamma_w^w$ and  $\Delta\gamma_{\alpha^*<\alpha}=\gamma_h^w-\gamma_h^h$; superscripts $h$ and $w$ tag for peptide hydrophobic $h$ or "waterphilic" $w$ zones; subscripts for the facing medium i.e. either hydrocarbonated lipid tails $h$ or water $w$). For sake of simplification let us assume that $\gamma_h^h=\gamma_w^w=0$ and $\gamma_w^h=\gamma_h^w=\gamma$. The interfacial energy per pore $E_{\text{int}}=\rho2\pi r\mathcal{E}_{\text{P}}$ is thus~:
\begin{equation}\label{Einterf}
E_{\text{int}}=2\pi r_{\text{P}}z \times \gamma \times \left\|{1-({r}/{r^*})}\right\|
\end{equation}
$E_{\text{int}}$ adds to the line tension in Eq.\ref{taupin}, increases $r_m^*$ and $E_m^*$, and  introduces a minimum for $E_m$ at $r^*<r_m^*$, allowing pores of energy smaller than $E_m^*$ to be stable.

\textit{- Electrical work}~: 
From our neutron reflectivity experiments we know that the electric field tends to orientate lipid head-groups of the cathodic-leaflet.
The corresponding potential electric energy is exactly balanced by an asymmetric tension of this leaflet of energy $E_{\text{leaf}}=(A/A_\text{L})\mu_\text{L}U\cos(\theta)/{z}$, with $\theta$ the angle of head-groups dipoles with the field. When a pore opens, this tension is relaxed by $E_{\text{leaf}}\pi r^2/A$.
This leads to a simple expression for the electric work of lipid dipoles coming with one pore opening~:
\begin{equation}\label{We}
W_{\text{L}}=-\left({ {\pi r^2 \mu_\text{L}\cos(\theta)}/{A_\text{L}z} }\right) U
\end{equation}
$W_{\text{L}}$ amounts to removing all the lipid dipoles in the area $\pi r^2$. Its analogous to take off the charges accumulated at the surface of the dielectric\,\cite{Glaser:1988vn}, $-U^2{\epsilon_m\pi r^2}/{2z}$, is always negligible compared to $W_{\text{L}}$ for usual voltages. Eq.\ref{We} is thus the dominant electrical contribution. 

\textit{- Entropy}~: 
Voltage clamp results oblige us to introduce entropy. The translational entropy of $N_\text{L}$ lipids in an area $A$ is $kN_\text{L}\ln\left({eA/N_\text{L}A_L}\right)$. For a held membrane under tension, the opening of a pore of area $\pi r^2$ 
only relaxes a bit the tension and lets the overall area unchanged.
So, the accessible area is reduced by the "excluded-area" of the pore. The entropy cost is~:
\begin{equation}\label{Slip}
-TS_{\text{L}}=kT{\pi r^2}/{A_\text{L}}
\end{equation}
As for adsorbed peptides, their translational entropy is reduced by this excluded-area but also by their localization in pores assumed immobile. The resulting cost per pore is~: $-TS_{\text{P}}=kTx\left({x/{x^*}}\right)\pi r^2$. 
Since $x/x^*\simeq 1$ and $x\ll 1/A_\text{L}$, thus $S_{\text{P}}\ll S_{\text{L}}$. It can be checked that it will be the same for all contributions of peptides to the entropy loss (rotational or conformational freedom etc). This is due to lipids outnumbering peptides and to the extensiveness of entropy. Eq.\ref{Slip} is thus the dominant contribution to the entropy cost of one pore.

Finally, the sum of Eq.\ref{taupin} and Eq.\ref{Emp} to \ref{Slip} estimates the free energy per pore. In particular, it accounts for the voltage-temperature dependence of pore opening. From Eq.\ref{We}-\ref{Slip}, at the transition $\partial U^*/\partial T^*=  k/\left({\mu_{\text{L}}\cos(\theta)/z}\right)$. With $\theta=77.4^\circ$\,\cite{Reigada:2011fk}, one obtains $\partial U^*/\partial T^*= 4.5\text{\,mV/K}$, in good agreement with the result in Fig.\ref{fig3}. Here, one understands that beyond an epiphenomenon, voltage effects amount to put the pore-opening transition in the correct temperature-window. 
Voltage-induced peptide pores always open with peptides adsorbed on the anodic-leaflet. 
In this letter, consistently with our experiments, we propose that this asymmetry, as well as the electric work, originate from the membrane rather than from the peptide. We also argue that the main contribution to the entropy cost comes from an "excluded-area" effect on lipid entropy.
This places the membrane in central position for the energetics of pore opening.
This "lipocentric view" has already been   suggested mainly in view of \textit{in silico} experiments\,\cite{Fuertes:2011fk}. Here, we attempt to formalize it in a more comprehensive way that could have a  general impact in modeling large molecules incorporation. 

\begin{acknowledgments}
\small{We thank F.~Cousin, L.T.~Lee, Y.~Tran for help in reflectivity experiments and samples preparation and G.~Brotons for encouraging discussions.}
\end{acknowledgments}

%

\end{document}